\pdfoutput=1
\documentclass{moriond}
\usepackage{xspace}
\def\PLB{{\em Phys. Lett.}  B }
\def\PRL{{\em Phys. Rev. Lett.} }
\def\JHEP#1#2#3{{\em JHEP} {\bf #1} (#2), #3}
\def\PRD{{\em Phys. Rev.} D }

\def\be{\begin{equation}}
\def\ee{\end{equation}}
\def\bea{\begin{eqnarray}}
\def\eea{\end{eqnarray}}

\def\TeV{\ensuremath{\mathrm{TeV}}\xspace}
\def\pt{\ensuremath{p_\mathrm{T}}\xspace}
\def\Dz{\ensuremath{\mathrm{D}^{0}}\xspace}
\def\Dp{\ensuremath{\mathrm{D}^{+}}\xspace}
\def\Ds{\ensuremath{\mathrm{D}_\mathrm{s}^{+}}\xspace}
\def\Lc{\ensuremath{\Lambda_\mathrm{c}^{+}}\xspace}
\def\LcD{\ensuremath{\Lc/\Dz}\xspace}
\def\pT{\pt}
\def\Xic{\ensuremath{\Xi_\mathrm{c}^{0,+}}\xspace}
\def\Xicz{\ensuremath{\Xi_\mathrm{c}^{0}}\xspace}

\def\Oc{\ensuremath{\Omega_\mathrm{c}^{0}}\xspace}
\def\Sigc{\ensuremath{\Sigma_\mathrm{c}^{0,++}}\xspace}
\def\epem{\ensuremath{\mathrm{e}^+\mathrm{e}^-}\xspace}
\def\ep{\ensuremath{\mathrm{e}^-\mathrm{p}}\xspace}
\def\pPb{\textrm{p--Pb}\xspace}
\def\PbPb{\textrm{Pb--Pb}\xspace}
\def\sqrts{\ensuremath{\sqrt{s}}\xspace}
\def\sqrtsNNfive{\ensuremath{\sqrt{s_\mathrm{NN}} = 5.02~\TeV}\xspace}
\def\sqrtsfive{\ensuremath{\sqrt{s} = 5.02~\TeV}\xspace}
\def\sqrtsthirt{\ensuremath{\sqrt{s} = 13~\TeV}\xspace}
\newcommand{\gevc}{\ensuremath{\mathrm{GeV}/c}\xspace}
\newcommand{\GeVc}{\gevc}

\newcommand{\figref}[1]{Fig.~\ref{#1}}

\newcommand{\Photo}{}
\graphicspath{{./figs/}}

\begin{document}
\vspace*{4cm}
\title{CHARM QUARK HADRONISATION STUDIES IN PP COLLISIONS WITH ALICE}

\author{J. WILKINSON, on behalf of the ALICE Collaboration}

\address{GSI Helmholtzzentrum f\"{u}r Schwerionenforschung GmbH,\\
Planckstra{\ss}e 1, 64291 Darmstadt, Germany }

\maketitle\abstracts{
In this contribution, the latest results for measurements of charm baryons in proton--proton collisions at $\sqrt{s}=5.02\mathrm{~and~} 13\,\mathrm{TeV}$ are presented. The production yields of $\Lc$, \Xicz, $\Oc$, and $\Sigc$ are shown along with their yield ratios to \Dz mesons, and observations about charm hadronisation with respect to previous results from lepton colliders are discussed. Further differential measurements are shown of the \LcD yield ratio as a function of charged-particle multiplicity. For the first time, the production of \Lc is presented in pp collisions down to $\pt = 0$, and the total charm production cross section and relative charm hadronisation fractions in \pPb collisions are computed.
}

\section{Introduction}

Heavy quarks (charm and beauty) have masses much larger than the characteristic energy scale of QCD interactions, $\Lambda_\mathrm{QCD}$. Due to this, they are typically produced in hard scattering processes with large $Q^2$, meaning that their production can be described well using perturbative QCD (pQCD) calculations. Typically the production of heavy-flavour hadrons is factorised into three main components: the parton distribution functions (PDFs) that describe the momentum distributions of quarks in the colliding hadrons; the partonic interaction cross sections to produce the heavy quarks; and the fragmentation functions, which describe the hadronisation of quarks to specific species. The non-perturbative fragmentation functions are determined from experiment in \epem and \ep collisions, and are assumed to universally define the hadronisation ratios of charm hadron species in other collision systems. Hadron-to-hadron production ratios are particularly sensitive to changes in the hadronisation mechanisms, since the contributions from the PDFs and scattering cross sections cancel out, making charm baryon measurements an important tool in testing the assumptions of the factorisation approach.

Recent measurements by the ALICE Collaboration of the production of D mesons showed that the strange and non-strange meson-to-meson yield ratios $\Ds/\Dz$ and $\Dp/\Dz$ are flat as a function of \pt~\cite{alice:Dmespp5}, meaning that there are no momentum-dependent modifications to charm meson hadronisation. The ratios are also in line with the expected values from previous \epem measurements. In contrast, the \Lc/\Dz yield ratios in pp and \pPb collisions at \sqrtsNNfive are significantly higher than the \epem measurements for $\pt<10\,\GeVc$ and exhibit a decreasing trend towards high transverse momentum~\cite{alice:LcpppPb5}. In addition, MC calculations tuned to hadronisation ratios from \epem collisions, such as PYTHIA with the Monash tune~\cite{monash}, are unable to reproduce the measurements in pp collisions. Instead, additional hadronisation mechanisms must be considered in order to describe the data. This implies that the previously assumed universality of fragmentation as the main hadronisation process in the factorisation approach does not hold true for baryon production. In order to fully characterise these effects in hadronic colliders, it is important to make precise measurements of further charm baryon states with the widest possible momentum coverage, as well as measuring differentially in terms of other observables such as the event multiplicity.

\section{Results}

\begin{figure}[h!tb] \centering
\begin{minipage}{0.42\linewidth}
\centerline{\includegraphics[width=0.9\linewidth]{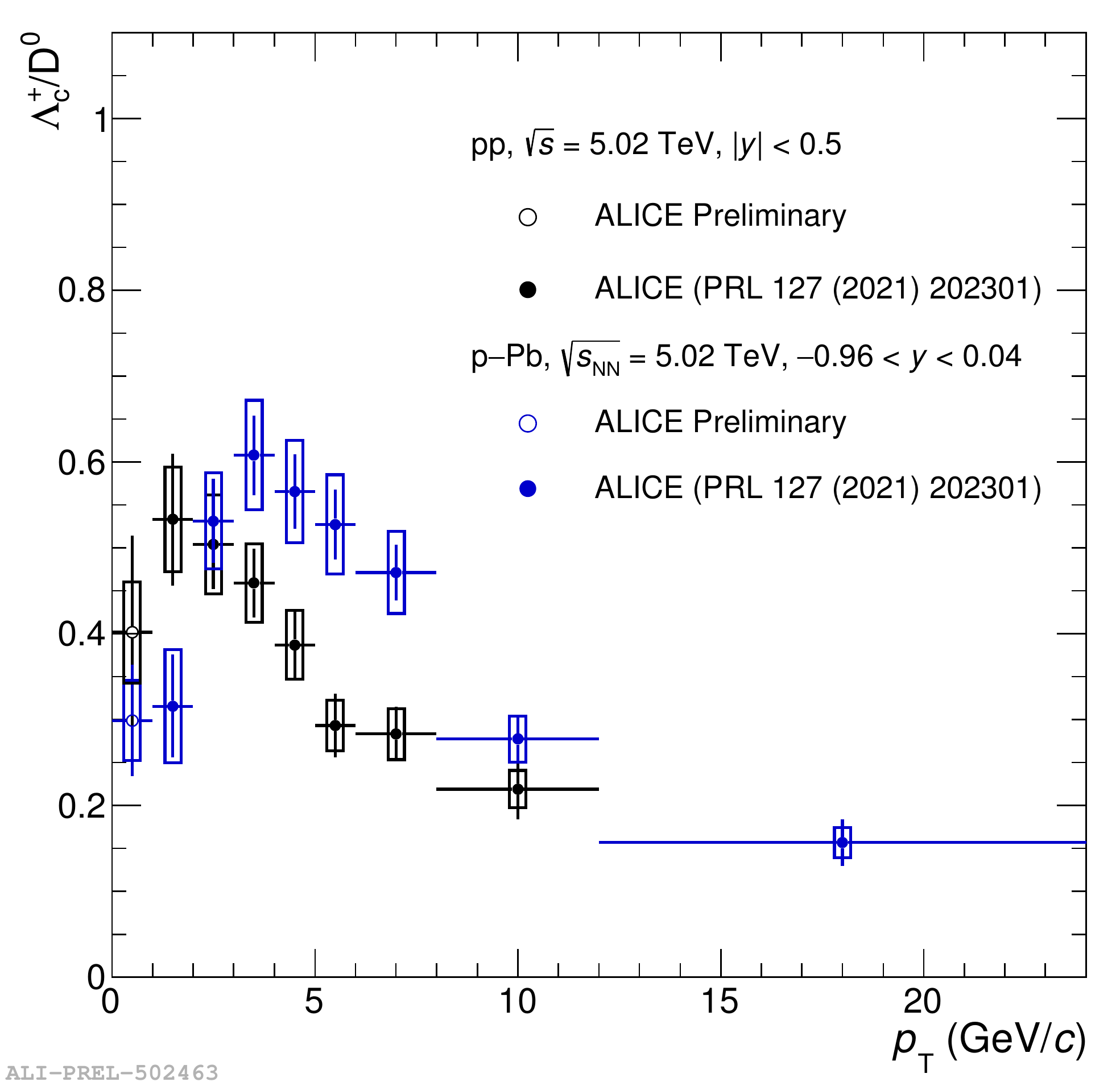}}
\end{minipage}
\begin{minipage}{0.42\linewidth}
\centerline{\includegraphics[width=0.9\linewidth]{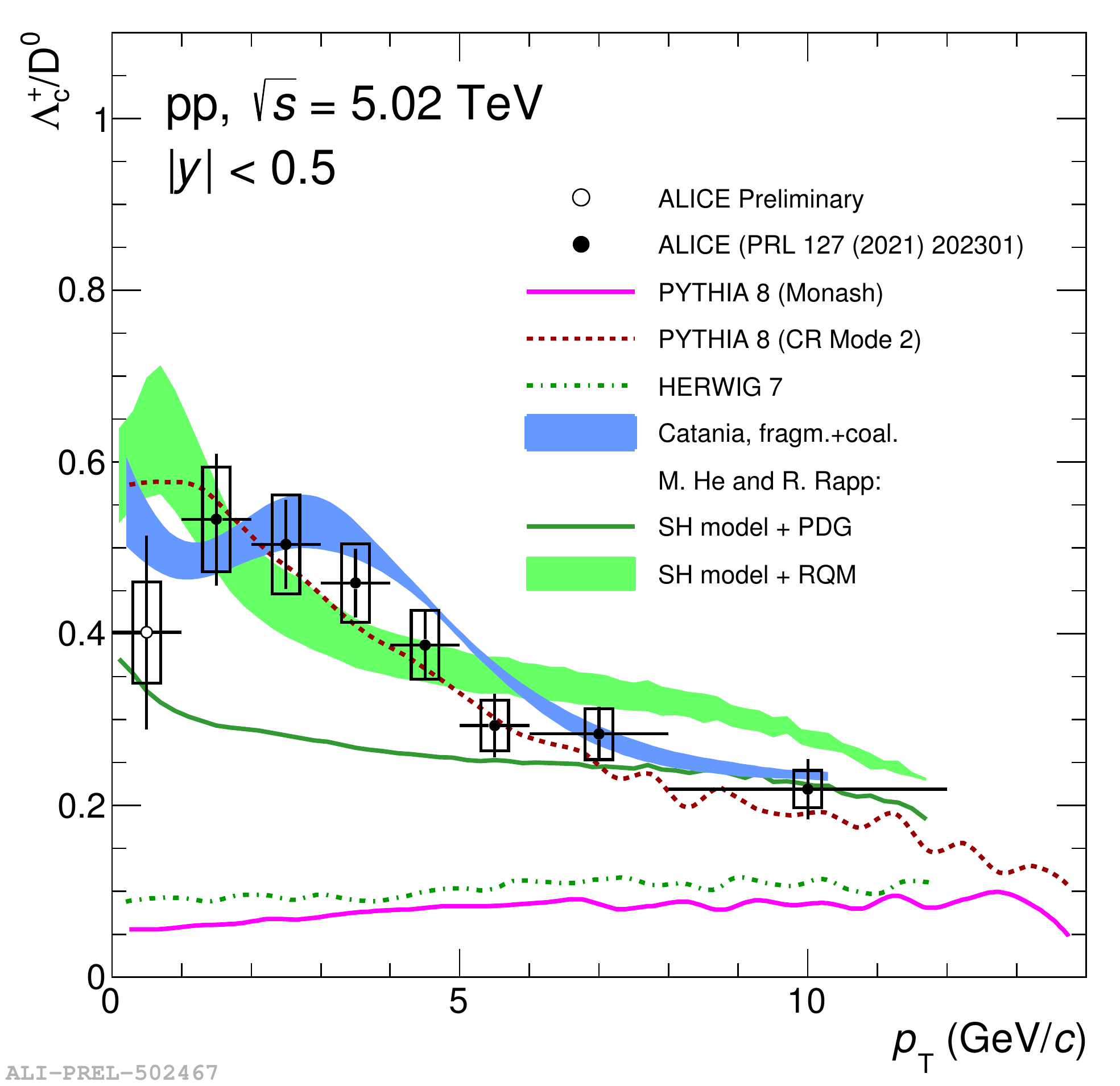}}
\end{minipage}
\caption[]{ALICE measurements of the \LcD baryon-to-meson yield ratio at \sqrtsNNfive. Left: Comparison of pp and \pPb collisions~\cite{alice:LcpppPb5}, including preliminary measurement of the interval $0<\pt<1\,\GeVc$. Right: Comparison between \LcD in pp collisions with models~\cite{monash,herwig,alice:LcpppPb5,pythCR,catania,rqm}.}
\label{fig:lclow}
\end{figure}

The baryon-to-meson yield ratio $\Lc/\Dz$ is presented down to $\pt=0$ in \pPb collisions at \sqrtsNNfive, and for the first time in pp collisions at \sqrtsfive, in \figref{fig:lclow} (left). In both collision systems, the measurements in the interval $0<\pt<1\,\GeVc$, indicated by the open circles, were performed in the channel $\Lc\to\mathrm{pK}^0_\mathrm{S}$, with candidates reconstructed from their decay tracks using the KFParticle package~\cite{kfparticle} and machine learning selections applied using the XGBoost gradient boosting algorithm~\cite{xgboost}. The measurements for $\pT>1\,\GeVc$ are from combined measurements of the decay channels $\Lc\to\mathrm{pK}^0_\mathrm{S}$ and $\Lc\to\mathrm{pK}\pi$ in both collision systems~\cite{alice:LcpppPb5}. A non-flat distribution is evident in both systems as a function of \pt, and there is a hint of a hardening in the \pt spectrum in \pPb collisions with respect to pp. In the right panel, the pp measurements are compared with model calculations. Models tuned to hadronisation ratios from \epem collisions, such as PYTHIA with the Monash tune~\cite{monash} and HERWIG~\cite{herwig} are unable to describe either the shape or the magnitude of the \LcD yield ratio in pp collisions. Models including additional hadronisation effects provide a better description of the results, for instance the Catania model~\cite{catania}, which considers quark coalescence in addition to fragmentation; PYTHIA calculations with enhanced colour reconnection beyond the leading order~\cite{pythCR}; and the SH + RQM model~\cite{rqm}, which is a statistical hadronisation approach including feed-down from yet-unmeasured resonant charm baryon states.

\begin{figure}[h!tb] \centering
\includegraphics[width=0.82\linewidth]{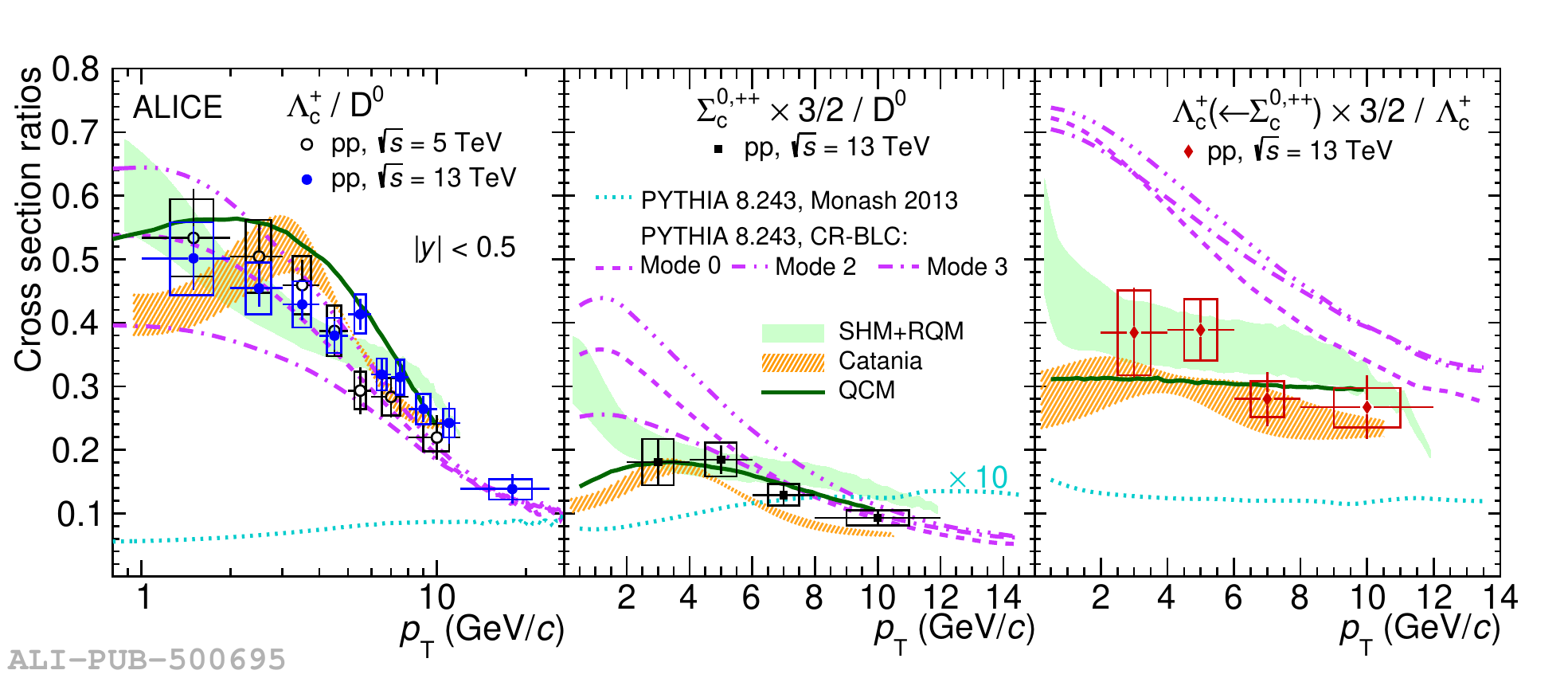}
\caption[]{Measurements of (left) \LcD, (middle) $\Sigc/\Dz$, and (right) contribution of \Sigc feed-down to the prompt \Lc cross section in pp collisions at \sqrtsthirt~\cite{alice:Sigc}.}
\label{fig:sigc}
\end{figure}

The \Sigc baryon states are potential contributors to the prompt cross section of \Lc baryons, as they decay strongly and are indistinguishable from prompt production. These states are expected to be suppressed with respect to \Lc due to the relative rarity of spin-1 diquark states produced in collisions. Measurements of \Sigc production are shown in~\figref{fig:sigc}, in particular the middle panel which shows the $\Sigc/\Dz$ production cross section, and the right panel which shows the feed-down contribution from \Sigc to \Lc~\cite{alice:Sigc}. A significant increase, by a factor of approximately 10, is seen in the $\Sigc/\Dz$ ratio with respect to the Monash tune of PYTHIA, while models with coalescence effects and statistical hadronisation with additional baryon states describe the data well. The fraction $\Lc(\leftarrow\Sigc)\times3/2$ is also significant, with a \pt-integrated value of $0.38\pm0.06\mathrm{(stat.)}\pm0.06\mathrm{(syst)}$. This implies that a sizable proportion of the measured prompt \Lc originates from the decays of heavier resonant states, and notably the PYTHIA predictions with enhanced colour reconnection appear to overestimate this contribution.

\begin{figure}[h!tb]\centering
\begin{minipage}{0.44\linewidth}
\centerline{\includegraphics[width=0.9\linewidth]{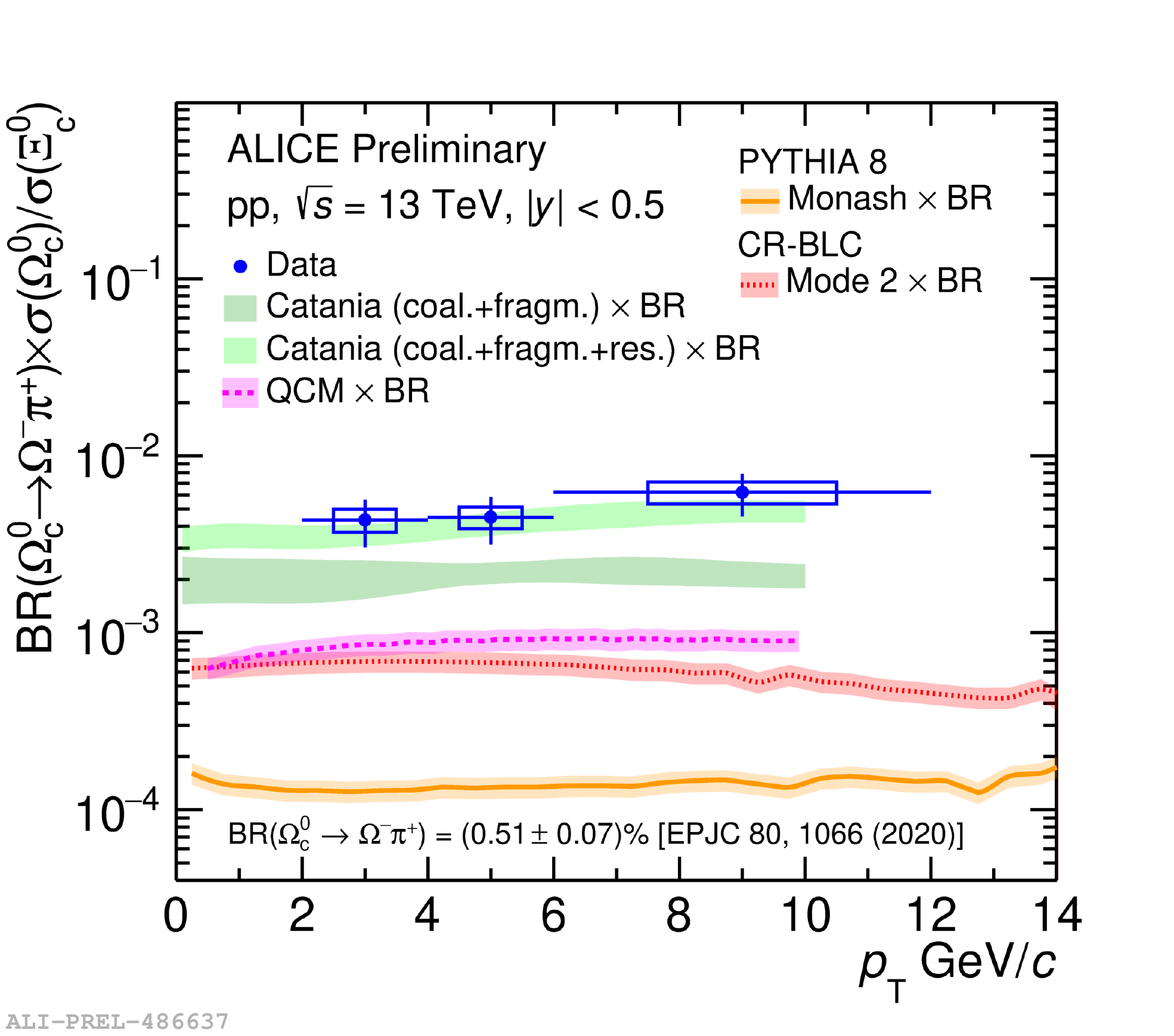}}
\end{minipage}
\begin{minipage}{0.44\linewidth}
\centerline{\includegraphics[width=0.85\linewidth]{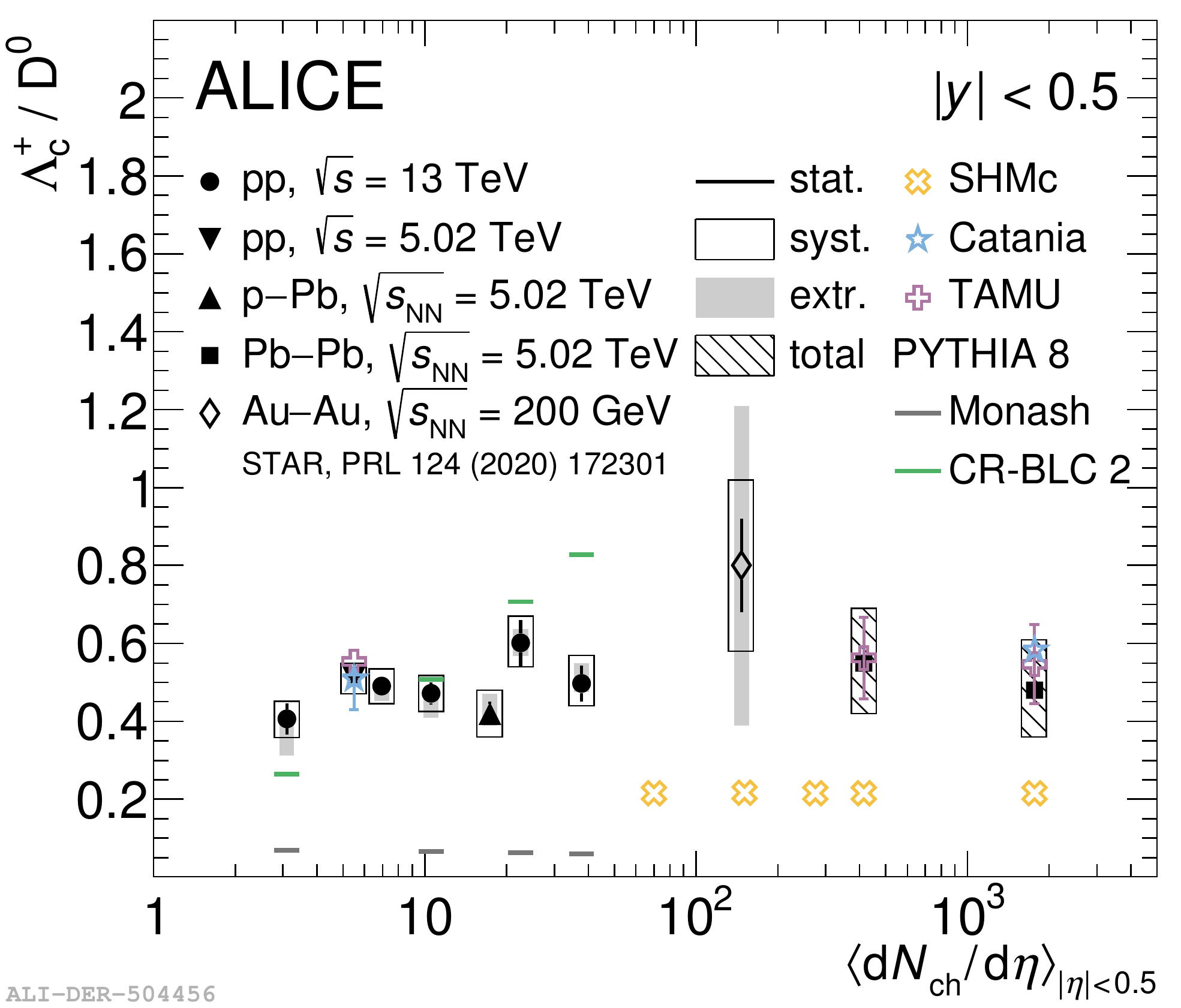}}
\end{minipage}
\caption[]{Left: $\mathrm{BR}(\Oc\to\Omega^-\pi^+)\times\sigma(\Oc)/\sigma(\Xicz)$ baryon-to-meson yield ratio in pp collisions at \sqrtsthirt. Right: \pt-integrated \LcD yield ratio as a function of charged-particle multiplicity, comparing results from pp, \pPb and \PbPb collisions in ALICE with Au--Au collisions at STAR.}
\label{fig:Oc}
\label{fig:LcDmult}
\end{figure}

The potential role of strangeness in the hadronisation processes may be considered by measuring \Xic and \Oc baryons. Both have been measured in pp collisions at \sqrtsthirt, and as an example the $\Oc/\Xicz$ ratio is shown in the left panel of~\figref{fig:Oc}. As the branching ratio of the decay channel $\Oc\to\Omega^-\pi^+$ is not yet experimentally known, the \Oc cross section is not corrected for the branching ratio. This baryon-to-meson yield ratio is underpredicted by two orders of magnitude by PYTHIA calculations using the Monash tune, and by one order of magnitude for PYTHIA including beyond-leading-order colour reconnection. The Catania model in this case comes closest to describing the data, but while the non-strange baryons were described well by the coalesence + fragmentation case, for $\Oc/\Xicz$ adding an extra contribution of feed-down from resonant baryon states better describes the data.

Differential measurements of baryon and meson production in terms of the charged-particle multiplicity are shown in~\figref{fig:LcDmult} (right). While a significant modification of the \pt distribution of the \LcD yield ratio has been observed between low- and high-multiplicity pp collisions at \sqrtsthirt~\cite{alice:LcvsMult}, the \pt-integrated \LcD yield ratio spanning three orders of magnitude in multiplicity show no significant dependence on the event multiplicity. This further implies that the collision system dependence of the overall \LcD yield ratio between leptonic and hadronic collisions is not simply related to the number of charged particles produced, but rather additional hadronisation processes on top of vacuum fragmentation must be considered for hadronic collisions.

\begin{figure}[h!tb]\centering
\begin{minipage}{0.43\linewidth}
\centerline{\includegraphics[width=0.9\linewidth]{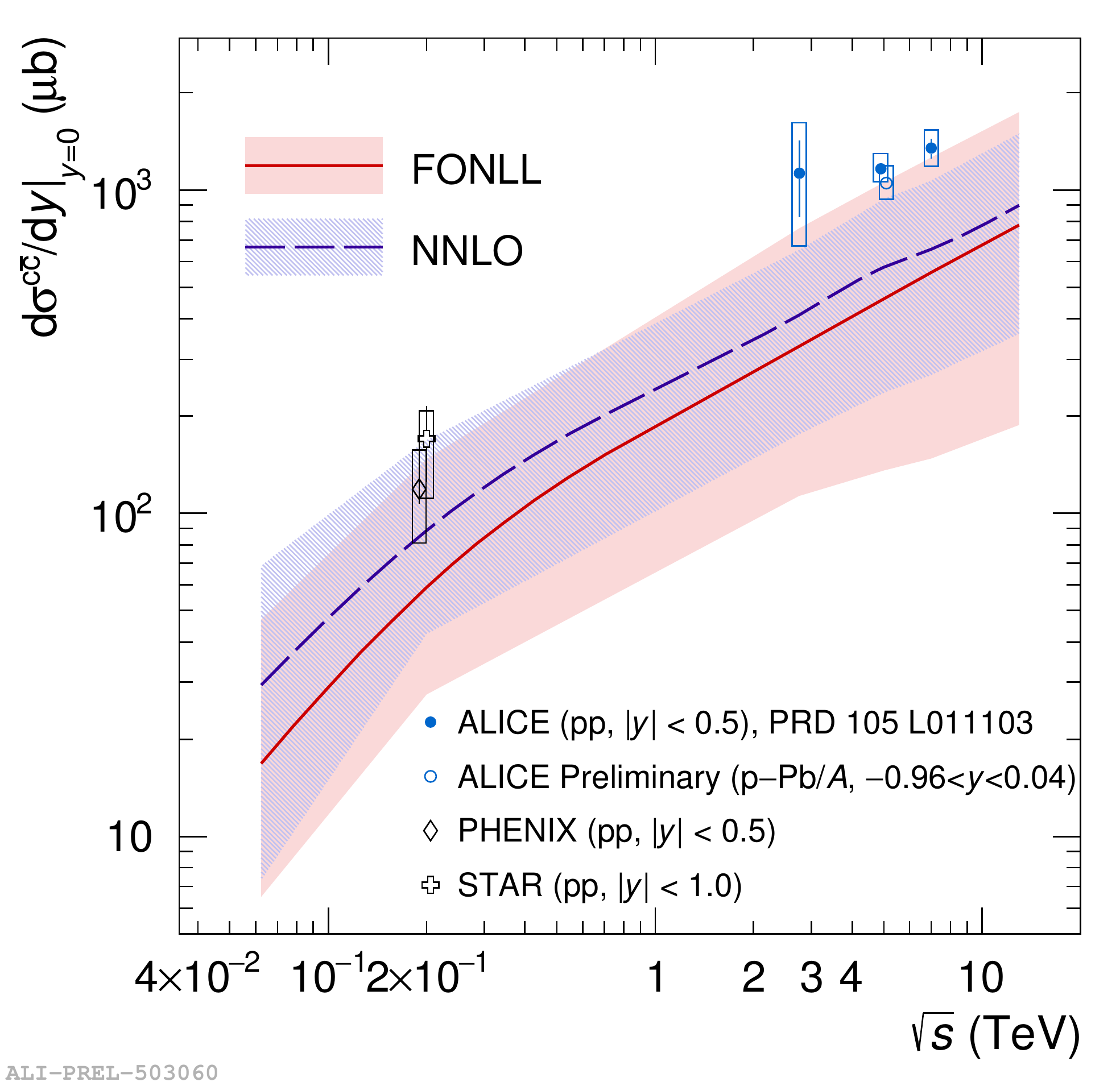}}
\end{minipage}
\begin{minipage}{0.43\linewidth}
\centerline{\includegraphics[width=0.9\linewidth]{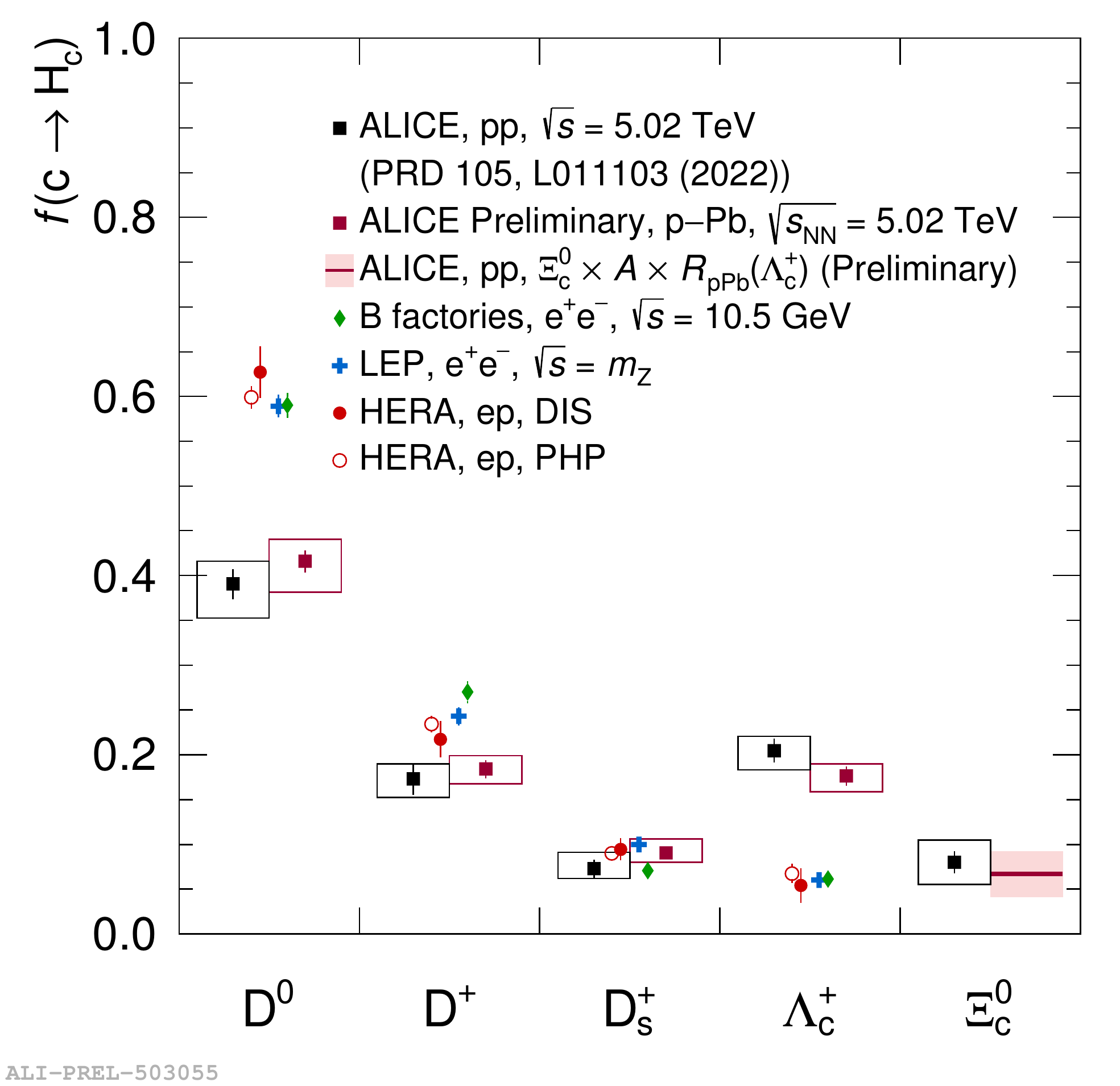}}
\end{minipage}
\caption[]{Left: Total charm production cross section at mid-rapidity in pp and p--Pb collisions as a function of collision energy. Right: Relative hadronisation fractions of ground-state charm hadron species measured by ALICE in pp and p--Pb collisions, compared with \epem and \ep collisions.}
\label{fig:crosssec}
\end{figure}

The measurement of all ground-state charm hadron species in a broad \pt range allows the total charm production cross section to be derived at mid-rapidity with minimal dependence on models, as shown in \figref{fig:crosssec} (left) as a function of collision energy. The total $\mathrm{c\bar{c}}$ cross section was computed for the first time in \pPb collisions at \sqrtsNNfive and shown as the open blue circle, along with previous ALICE measurements~\cite{alice:ppfrag} in pp collisions at $\sqrts=2.76, 5.02,$ and $7\,\TeV$ shown as solid points. The results are compared with pQCD calculations under the FONLL and NNLO schemes, and the ALICE measurements are consistently seen to lie at the upper edge of the theory uncertainty bands. The cross section can also be split into the individual contributions from hadron species to give the relative hadronisation fractions, $f(\mathrm{c}\to\mathrm{H}_\mathrm{c})$, including the first measurement of the $f(\Xicz)$. The measurements in pp collisions~\cite{alice:ppfrag} and preliminary measurements in \pPb collisions are shown in the right panel of~\figref{fig:crosssec}, in comparison with measurements in \epem and \ep collisions. The two hadronic collision systems are consistent with one another, but a significant enhancement of \Lc and depletion of \Dz production are seen with respect to leptonic collisions.

Looking to the upcoming Run 3 of the LHC and beyond to ALICE 3, the higher rates of data taking and upgrades to detector systems will allow the precision and \pt coverage of charm baryon measurements to be greatly improved as well as experimental access to new species, e.g. multi-charm baryons such as $\Xi_\mathrm{cc}^{++}$.

\end{document}